\documentclass{blois}

\usepackage{amsmath}

\newcommand{\ttbar}{\ensuremath{\mathrm{t} \bar{\mathrm{t}}}}
\newcommand{\ttg}{\ensuremath{\ttbar \gamma}}
\newcommand{\ttz}{\ensuremath{\ttbar \mathrm{Z}}}
\newcommand{\tzq}{\ensuremath{\mathrm{t} \mathrm{Z} \mathrm{q}}}
\newcommand{\tttt}{\ensuremath{\ttbar\ttbar}}
\newcommand{\ttH}{\ensuremath{\ttbar\mathrm{H}}}
\newcommand{\pt}{\ensuremath{p_\mathrm{T}}}

\def\Journal#1#2#3#4{{#1} {\bf #2}, #3 (#4)}

\def\JINST{\em JINST}
\def\JHEP{\em JHEP}
\def\EPJC{{\em Eur. Phys. J.} C}

\newcommand{\Photo}{\includegraphics[height=35mm]{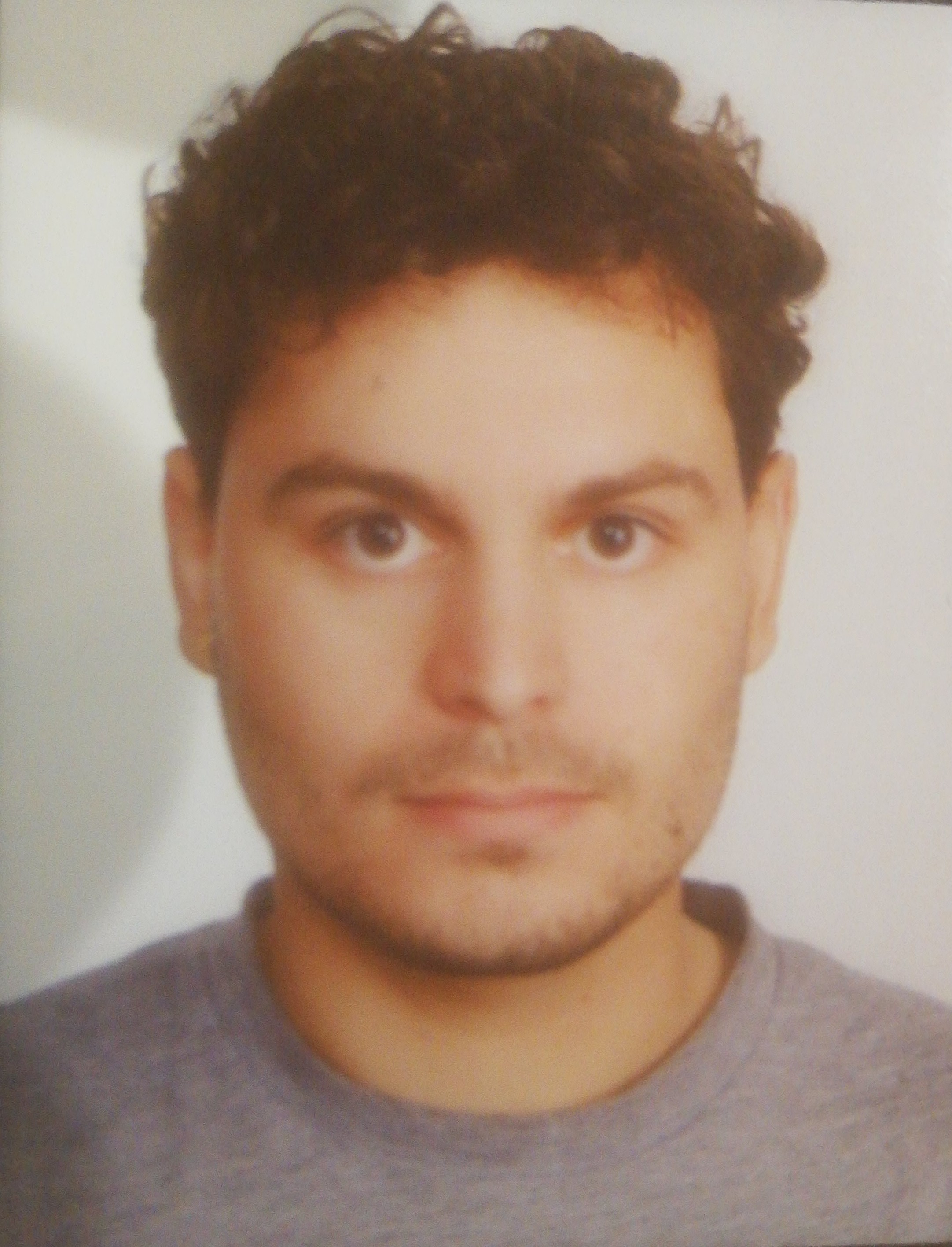}}

\begin{document}
\vspace*{4cm}
\title{RARE TOP QUARK PRODUCTION IN CMS}

\author{David L. Walter on behalf of the CMS Collaboration}

\address{Deutsches Elektronen-Synchrotron (DESY), Notkestraße 85,\\
22607 Hamburg, Germany}

\maketitle\abstracts{
Latest results from the CMS experiment at the LHC on top quark production in association with a Z boson or a photon, and the production of four top quarks, are summarized. Proton-proton collision data corresponding to an integrated luminosity of up to 138\,fb$^{-1}$, collected at 13\,TeV center-of-mass energy are used. Among the measured results, most precise inclusive cross sections, most stringent limits on models beyond the standard model and first differential measurements are presented. 
Overall, good agreement with standard model predictions is observed.}

\section{Introduction}
During Run-II of the CERN LHC in the years from 2016 to 2018, an unprecedented amount of proton-proton collision data at $\sqrt{s}=13$\,TeV corresponding to an integrated luminosity of 138\,fb$^{-1}$ has been collected by the CMS detector~\cite{CMS}. 
This allows for the first time detailed studies on rare top quark processes, including top quark pair production in association with a photon (\ttg) or Z boson (\tzq), single top quark production in association with a Z boson (\tzq), and the production of four top quarks (\tttt). 
These processes are of special interest since they contain electroweak couplings already at tree level and probe the standard model (SM) and beyond SM (BSM) physics like effective field theory (EFT), in a unique way. 
Furthermore they are important backgrounds in other analyses for example in measurements of \ttH\ production. 
The \tttt, even though it has the smallest expected cross section among the studied processes, is particular interesting since it is sensitive to the top-Yukawa coupling of the Higgs boson.
The presented analyses comprise updated inclusive cross sections and first differential measurements. 
Interpretations in the EFT framework and other BSM models are performed for the search for new physics.

\section{Top quark pair production in association with a photon}
The measurement of \ttg\ is one of the best ways to study the t-$\gamma$ coupling that appears already at tree level in this process. 
Two analyses focusing the single and double lepton (e or $\mu$) final states and using full Run-II data have been published~\cite{ttg_1l,ttg_2l}. 
In both measurements, the signal is defined in fiducial phase spaces based on particle level objects where the $\gamma$ is radiated from the initial state, final state, top quark, or one of the top quark decay products before hadronization. 
Processes with a misidentified $\gamma$ or where the $\gamma$ arises from a hadron are estimated using control samples in data. 
In the single (double) lepton analysis, events are selected that have exactly one (two) leptons, exactly one isolated photon and at least three (one) jets from which at least one has to be b tagged. In the dilepton analysis the two leptons are required to be of opposite charge and have a mass incompatible with the one of the Z boson.
The single lepton analysis suffers from much larger background contribution and control samples in data are used to estimate the contribution from events with nonprompt leptons as well. 
In both analyses control regions for processes with a Z boson and $\gamma$ are included to constrain the normalization of this background. 
The inclusive cross sections are measured to be $800\pm 7\text{ (stat) } \pm 46 \text{ (syst) } \mathrm{fb}$ and $174.4\pm 2.5\text{ (stat) } \pm 6.1 \text{ (syst) } \mathrm{fb}$ in the single and double lepton final state, respectively. 
With a precision of $5.8\%$ and $3.8\%$, the values are more precise than SM calculations at next-to-LO (NLO) and agree with them. 
Leading systematic uncertainties come from the integrated luminosity, parton shower modeling, and lepton selection and in the single lepton measurement also the jet energy scale and background normalization. 

\begin{figure}[!hbtp]
\centering
    \includegraphics[width=0.26\textwidth]{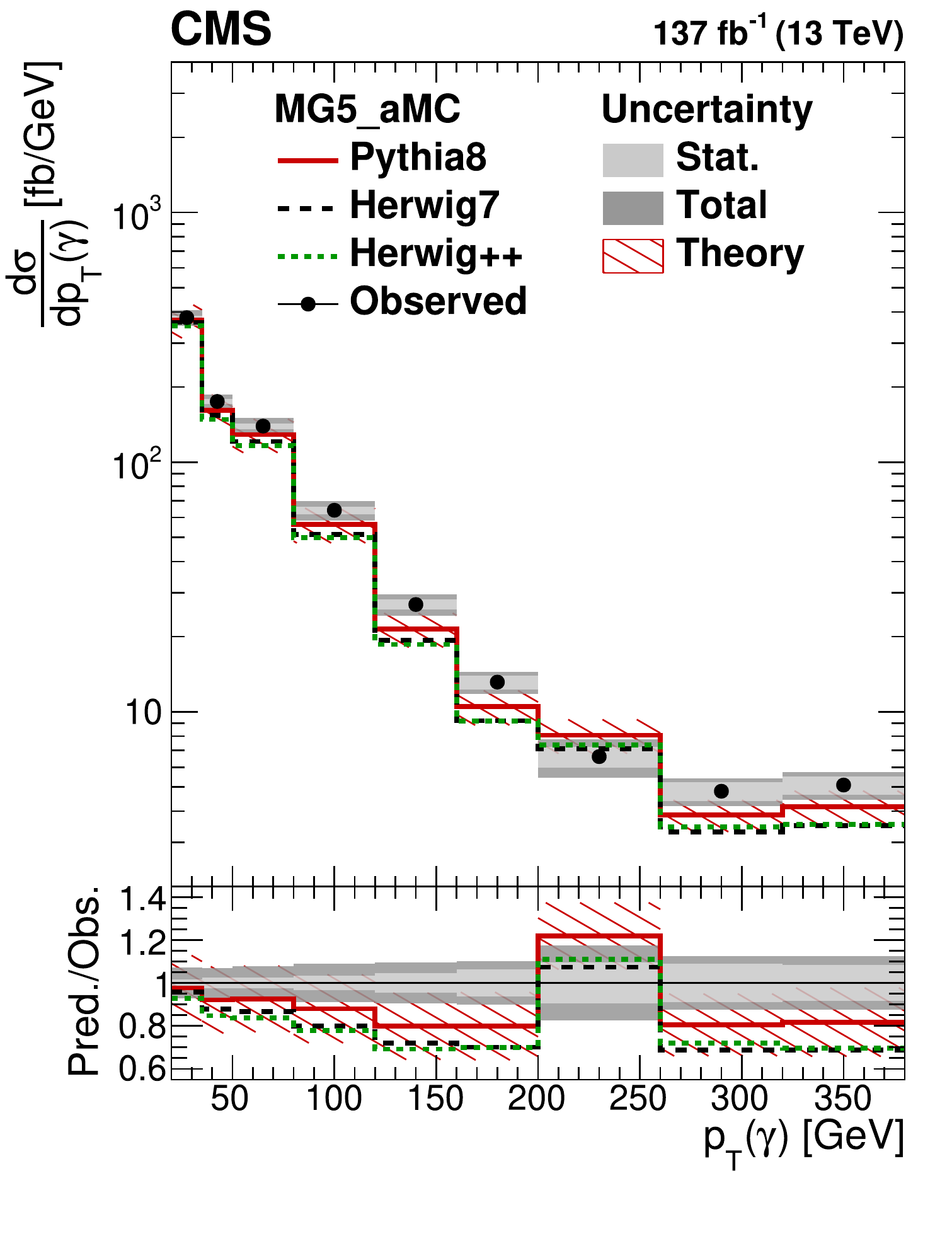}
    \includegraphics[width=0.26\textwidth]{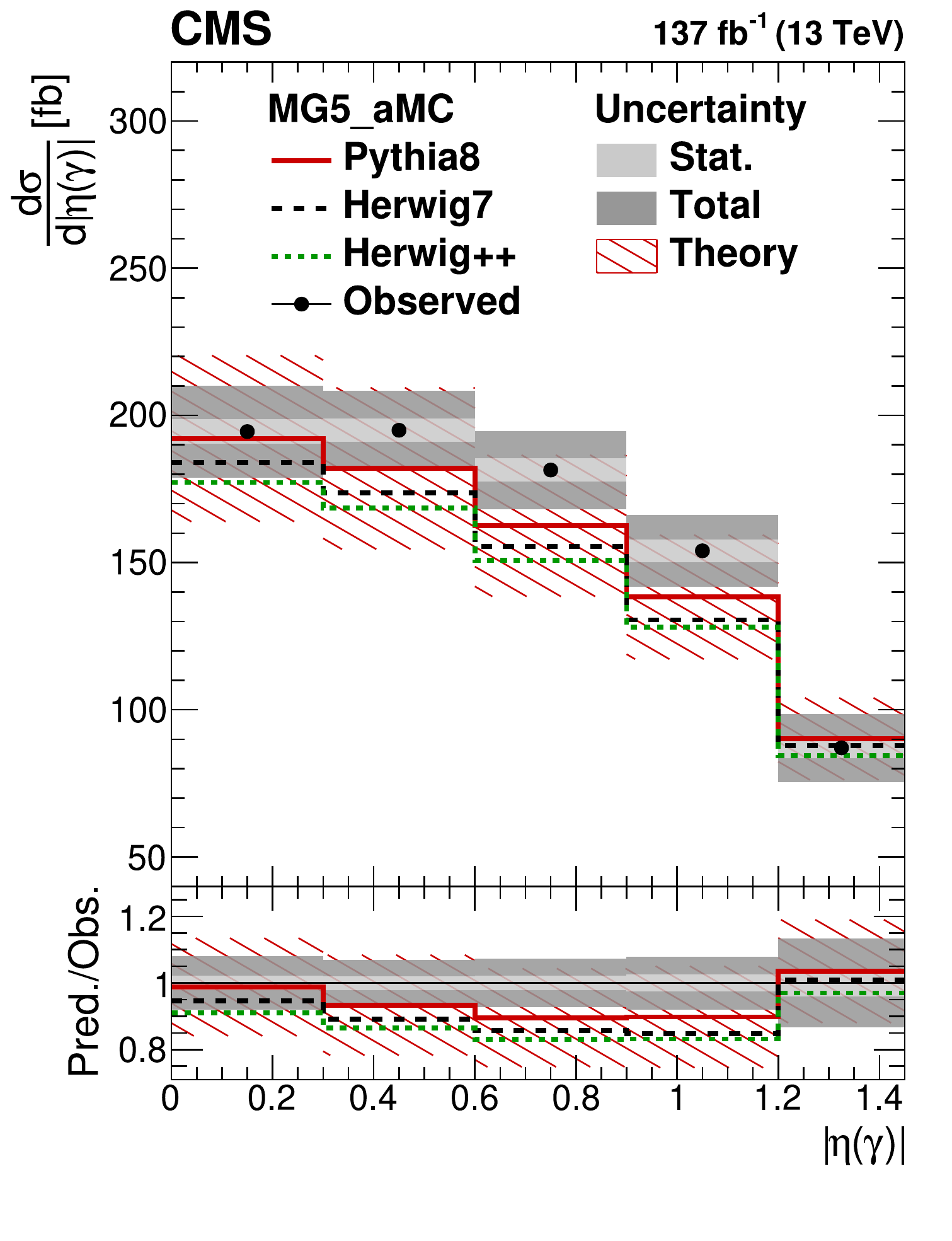}
    \includegraphics[width=0.45\textwidth]{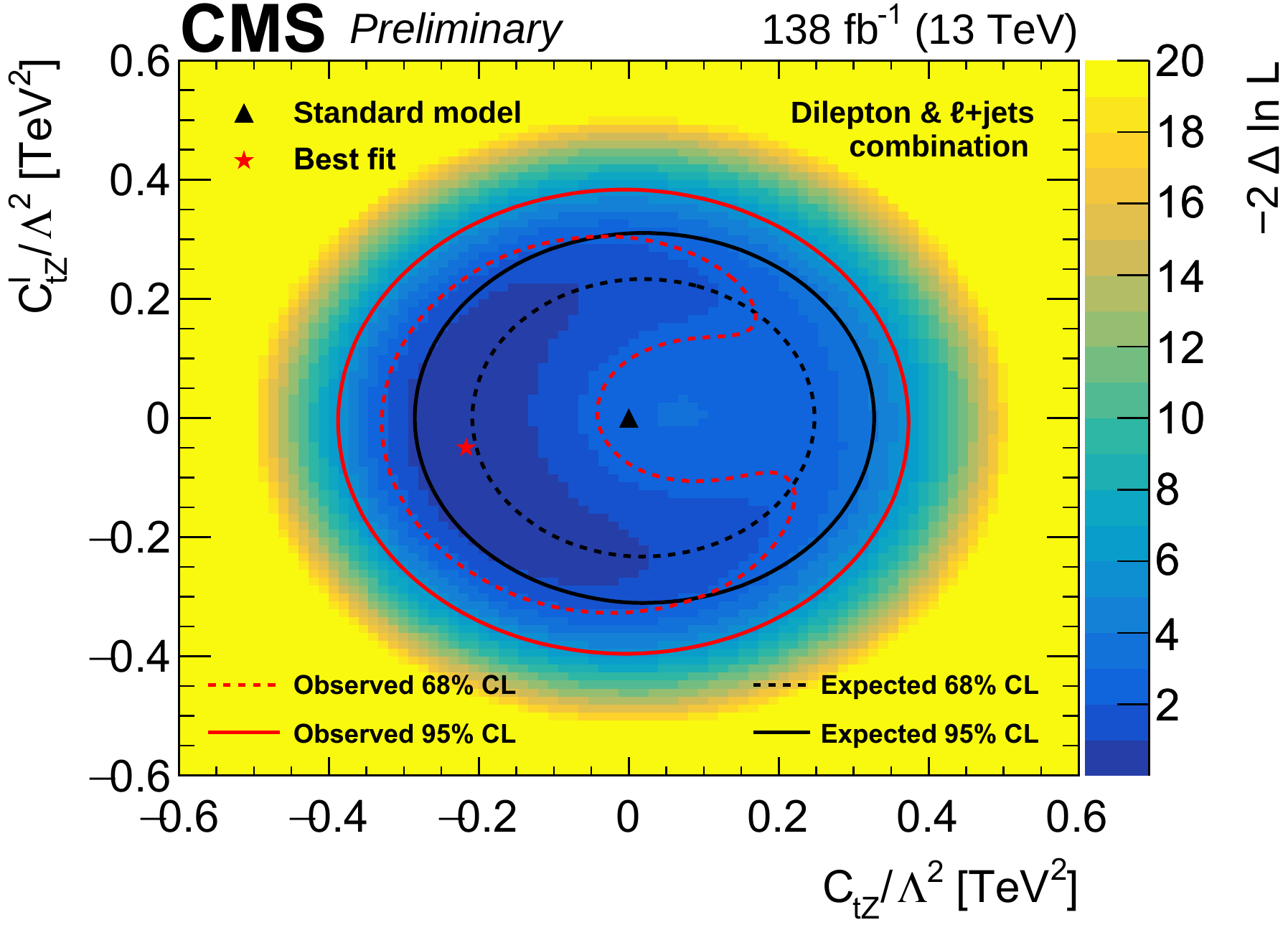}
    \caption[]{
    Left: Unfolded distribution of \pt($\gamma$) and $|\eta(\gamma)|$ at particle level from the single lepton analysis. The data (points) are compared to simulation with different parton shower models (colored lines). The lower panel shows the ratio of each simulation to the observed values~\cite{ttg_1l}.
    Right: The 2D likelihood of the EFT measurement of the combined single and double lepton measurements as a function of the Wilson coefficients $c_\mathrm{tZ}$ and $c_\mathrm{tZ}^\mathrm{I}$. The red dot shows the minimum of the likelihood while the red dotted and solid lines show the one and two sigma confidence intervals~\cite{ttg_2l}.
    }
\label{fig:ttg}
\end{figure}

In both analyses, the cross section is measured differentially as a function of kinematic variables of the $\gamma$ and the angular separation between the lepton and the $\gamma$. 
In the dilepton final state, additional variables include kinematic distributions of the leptons, dilepton pair, leading jet, and angles between these objects. 
To extract the signal, the background is subtracted from the data and the distribution is unfolded to particle level using the inverse response matrix, constructed from NLO simulation. 
The unfolded distributions are compared to simulation with different parton shower models and found to be overall in good agreement. 
Results from the single lepton analysis are shown for the transverse momentum (\pt) and pseudorapidity ($\eta$) of the $\gamma$ on the left side of Fig.~\ref{fig:ttg}. 

Both analyses are combined in an EFT interpretation to probe electroweak dipole moments. 
Therefore, simulated samples are reweighted and parameterized as function of the Wilson coefficients $c_\mathrm{tZ}$ and $c_\mathrm{tZ}^\mathrm{I}$. 
The events are binned in the \pt\ of the $\gamma$ and the likelihood scan is performed in 1D and 2D to extract limits on the Wilson coefficients. 
The 2D likelihood is shown on the right side of Fig.~\ref{fig:ttg}. 
The extracted limits are in agreement with SM calculations and the most stringent to date.

\section{Top quark pair production in association with a Z boson}
The \ttz\ production was studied in the final state with three or four prompt leptons (e or $\mu$) using only a partial dataset corresponding to 77.5\,fb$^{-1}$. 
A dedicated multivariate analysis (MVA) technique is used to discriminate prompt from nonprompt leptons while events containing the latter are estimated from control samples in data. 
Events are binned in lepton, jet, and b-jet multiplicities such that some of the bins have high purity in background processes with two massive vector bosons (WZ and ZZ) and serve to constrain their normalization. 
The inclusive cross section is measured to be $0.95\pm 0.05 \text{ (stat) } \pm 0.06\text{ (syst) pb}$ in agreement with the dedicated theory prediction of $0.86^{+0.07}_{-0.08} \text{ (scale) } \pm 0.03 \text{ (PDF) pb}$
including NLO in QCD and EW, and NNLL terms~\cite{ttZ_Theory}. 
Leading uncertainties on the measurement come from the WZ normalization and lepton identification. 


The analysis presents also the first differential measurements of this process where the \pt\ and an angular variable of the Z boson are measured using the inverse response matrix at parton level. 
The differential distributions are compared to predictions at NLO and next-to-NLO~\cite{ttZ_diff_Theory} precision and show a slight preference for the latter one.
An interpretation in the EFT framework and in the framework of anomalous couplings is performed in a similar way as done for the \ttg\ analyses.

\section{Single top quark production in association with a Z boson}
The study of \tzq\ is complementary to \ttz\ since the top is electroweak produced and also contains triple gauge boson couplings at LO. 
Measurements focusing the final state with three leptons using full Run-II data have been performed~\cite{tzq}. 
Prompt leptons are identified similar to the \ttz\ measurement and the nonprompt lepton background is estimated from control samples in data. 
The challenge of this process is to isolate the signal from various background processes, where event MVAs are trained that use characteristic variables e.g. related to the recoiling nature of the light flavor jet in respect to the top quark.
Additional control regions are defined to constrain the background normalization. 
The inclusive cross section is measured to be $87.9^{+7.5}_{-7.3} \text{ (stat) }^{+7.3}_{-6.0} \text{ (syst) }$fb.

\begin{figure}[!hbtp]
\centering
    \includegraphics[width=0.39\textwidth]{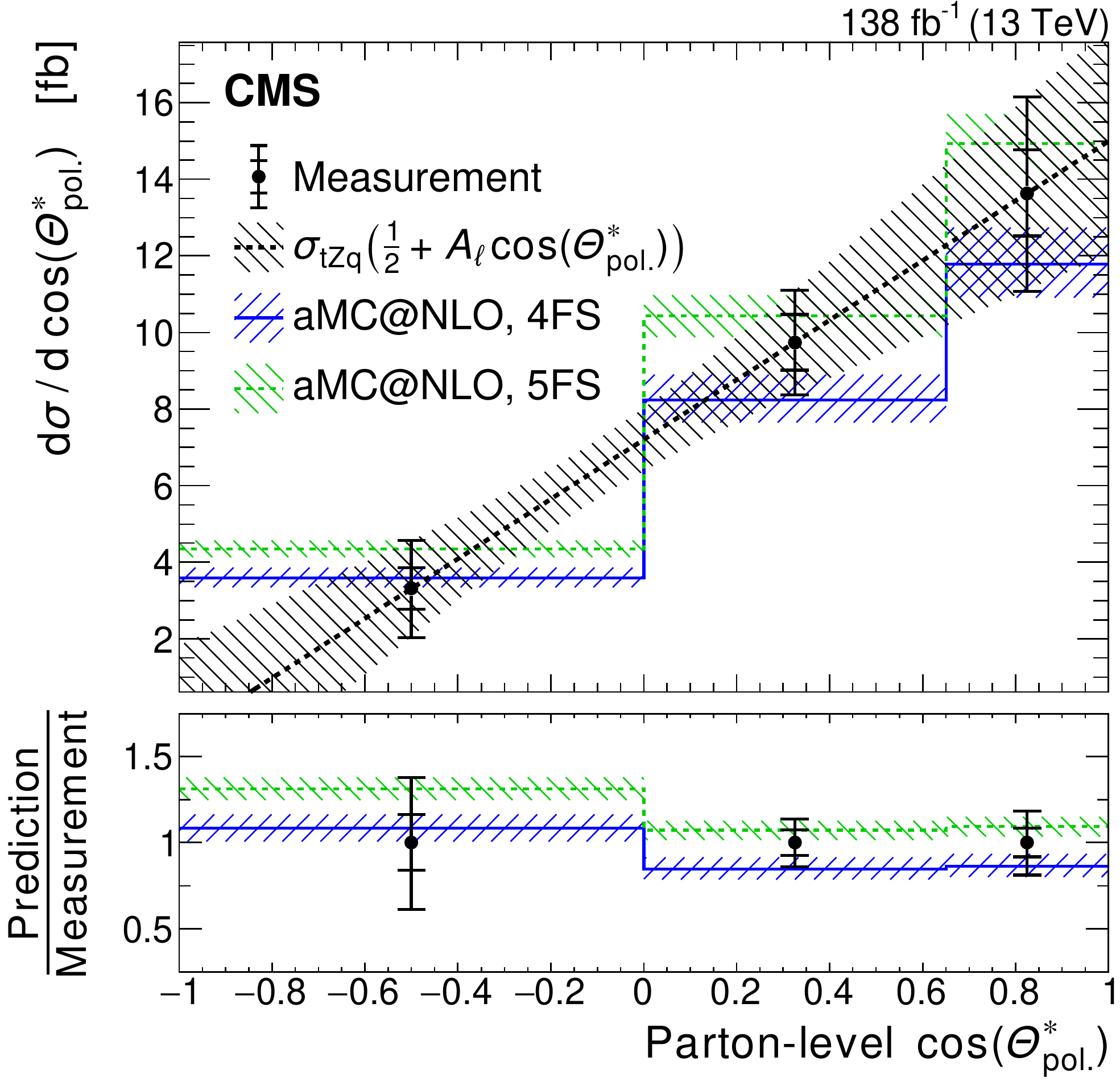}
    \includegraphics[width=0.595\textwidth]{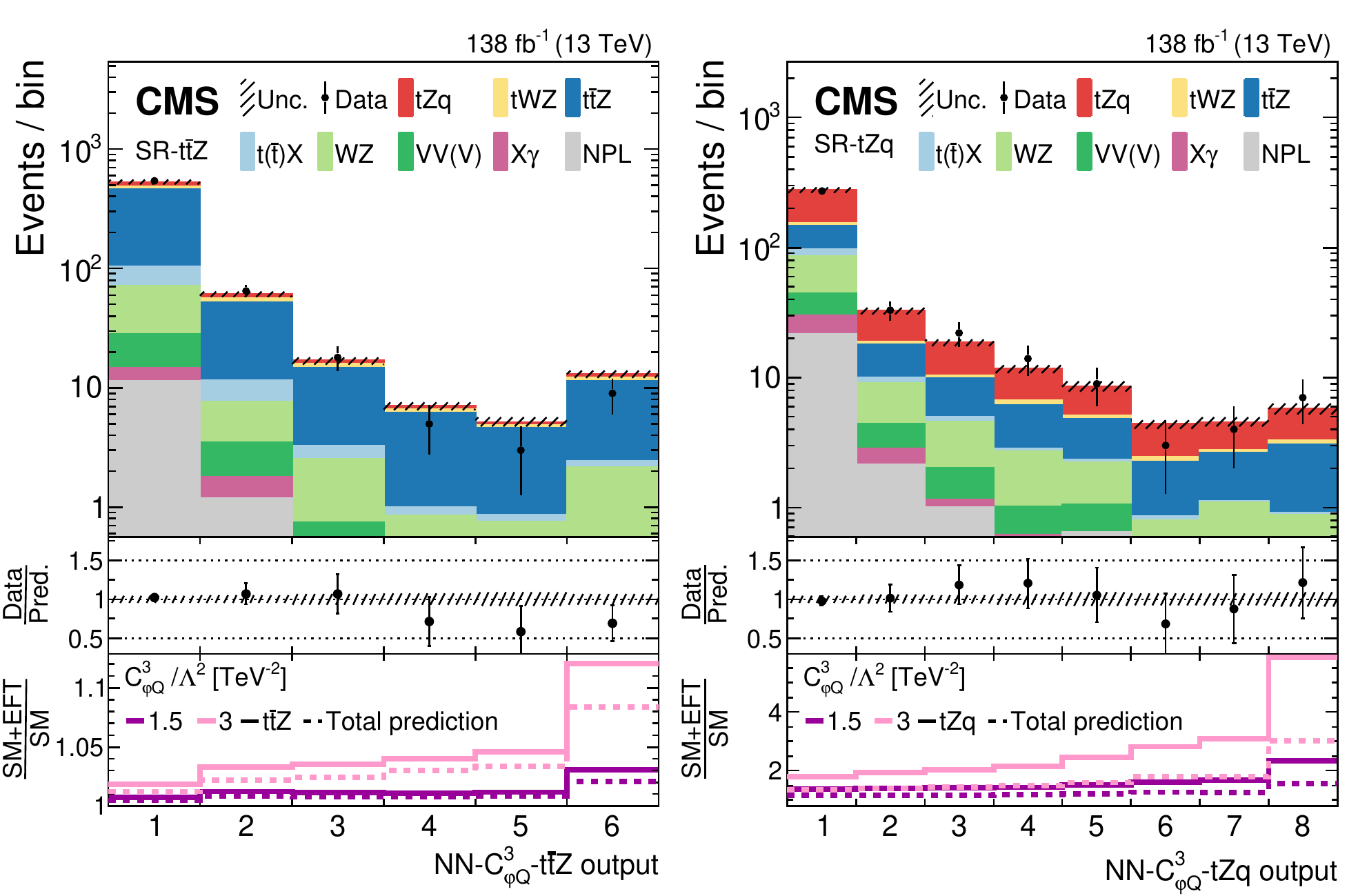}
    \caption[]{Left: Differential cross section of \tzq\ at parton level as a function of $\mathrm{cos}\left(\Theta^*_\mathrm{pol}\right)$ extracted from the measurement of the top quark spin asymmetry. The parameterization of the parton level bins is illustrated with the black dashed line.~\cite{tzq} Right: The \ttz\ and \tzq\ enriched regions for the extraction of 1D limits of an Wilson coefficient in the EFT analysis. In the middle panel the ratio of the data to the prediction is shown. The lower panel shows the expected increase in the number of \ttz\ or \tzq\ (solid) and total (dashed) events of the contribution for two given EFT scenarios.~\cite{EFT}}
\label{fig:tzq}
\end{figure}

For the first time in \tzq, differential cross sections are measured as function of various kinematic variables. 
A multidimensional likelihood fit is performed on different contributions of \tzq\ corresponding to the particle or parton level bins, including all backgrounds and systematic uncertainties. 
The signal region is binned in the observable and the event MVA.
In a similar way, the top quark spin asymmetry, proportional to the top quark polarization, is measured from an angular differential distribution, $\mathrm{cos}\left(\Theta^*_\mathrm{pol}\right)$, at parton level. The fit is therefore reparametrized as function of the spin asymmetry which is then directly extracted from the likelihood. 
The relation of the fitted top quark spin asymmetry is illustrated on the left side of Fig~\ref{fig:tzq}.
Results are compared to NLO SM calculations using the four and five flavor scheme for the parton distribution functions. 
Overall, they show good agreement with both predictions.

A separate analysis is focused on the EFT interpretation of the \tzq\ and \ttz\ sector in final states with three or four leptons~\cite{EFT}. 
In a first step, a multiclass neural network (NN) is used to separate different SM processes \tzq, \ttz, and WZ and enrich them in separate regions. 
In a second step, another NN is trained to separate the SM scenario from an EFT scenario.
Five different Wilson coefficients are tested, each one at a time and all five together, corresponding to six separate NNs for this second step. 
These NNs are used to bin the events in the enriched \tzq\ and \ttz\ regions as illustrated on the right side of Fig.~\ref{fig:tzq}. 
Likelihood fits are performed to extract 1D and 5D limits on the Wilson coefficients. 

\section{Production of four top quarks}
The production of four top quarks is of special interest as it is affected by the top-Yukawa coupling of the Higgs boson. 
A measurement is performed using events with two same sign or three leptons~\cite{tttt}.
Moreover, events are required to have large amount of hadronic activity, missing transverse energy and at least two b tagged jets to account for the four top quarks. 
Control samples in data are used to estimate backgrounds with nonprompt leptons or electron with misidentified charge.
A boosted decision tree is trained to separate the signal from the backgrounds. The significance of \tttt\ production is measured to be 2.7 standard deviations with respect to the background only hypothesis. 
While the measurement is statistically limited, also the modeling of background processes with additional b jets and the uncertainty on the jet energy scale have significant impacts.
The measurement is repeated and the contribution of the \ttH\ background is varied to extract a limit for the top-Yukawa coupling of 1.7 times the value expected from the SM. 
Furthermore, the measurement is interpreted in different BSM models where everything found to agree with SM expectations and limits have been set.



\newcommand{\arxiv}[1]{ArXiv:#1}
\newcommand{\cds}[1]{CDS:#1}
\newcommand{\doi}[2]{\mbox{#2}}
\newcommand{\pas}[1]{CMS-PAS-#1}

\end{document}